\documentclass{pazh_engl}
\usepackage{epsf}
\usepackage{flushrt}
\usepackage{graphicx}

\begin{document}

\sloppypar

\title{\bf RXTE observations of X-ray transients IGRJ17091-3624 and IGRJ18539+0727}

\author{\copyright 2003 A.A. Lutovinov \inst{1,2}, M.G.Revnivtsev\inst{2,1}}

\institute{Space Research Institute, Moscow, Russia
\and
Max-Planck-Institut f\"ur Astrophysik, Garching, Germany}

\authorrunning{LUTOVINOV \& REVNIVTSEV}
\titlerunning{}
\date{June 9, 2003}
\abstract{We present results of analysis of observations of two transient
sources IGR J17091-3624 and IGR J18539+0727 in April 2003 with RXTE 
observatory. Obtained energy spectra of sources, and also power spectra 
of their flux variations give us a possibility to classify them as X-ray 
binary systems in low/hard spectral state. Parameters of power spectrum
of IGR J18539+0727 indicates that this source is a black hole candidate.
}
\titlerunning{RXTE observations of IGR J17091-3624 and IGR J18539+0727}
\maketitle

\section*{Introduction}

X-ray sources IGR J17091-3624 and IGR J18539+0727 were discovered by
INTEGRAl observatory during deep survey of the Galactic central 
radian and routine scans of the Galactic plane in April 2003 
(Kuulkers et al. 2003, Lutovinov et al. 2003a).

The first source was detected on April 14, 2003 with IBIS telescope
aboard INTEGRAL observatory in the energy band 40-100 keV.
The source flux was measured to be 20 mCrab. In softer energy band 15-40 keV 
only upper limit on the source flux was obtained at that time -- 
$<$10 mCrab. Following observations of this source by INTEGRAL showed
that the source flux increased up to 40 and 25 mCrab in energy bands 15-40 and 40-100 keV respectively, that indicated that the spectrum of the source
probably softened (Kuulkers et al. 2003)

Analysis of archived data of the TTM telescope of the ``Roentgen'' observatory
aboard KVANT module of the MIR space station and WFC telescope of the BeppoSAX
observatory revealed sporadic appearances of this source during period
1994--2001 (Revnivtsev et al. 2003, in't Zand et al. 2003). 
Observations of the source in radio band, performed by VLA telescopes 
after the source discovery in April 2003, revealed the presence
of possible radio companion (Rupen et al. 2003).

After several days, on April 17-18, 2003, during the Galactic plane scan
and observations of microquasar GRS 1915+105 the second source was 
discovered. The source flux was measured to be 20 mCrab in energy bands
15-40 and 40-100 keV (Lutovinov et al. 2003a).

In order to try to understand the nature of the newly discovered sources
almost immediately after INTEGRAL discovery the RXTE observations were
performed. RXTE observatory have high sensitivity in the energy band 3-20 
keV and can be successfully used for spectral and timing analysis of the
sources. In this paper we present the results of analysis of these observations.

\section*{Observations, data analysis and results}

\begin{figure*}[htb]
\hbox{
\includegraphics[width=\columnwidth,bb=55 410 550 720,clip]{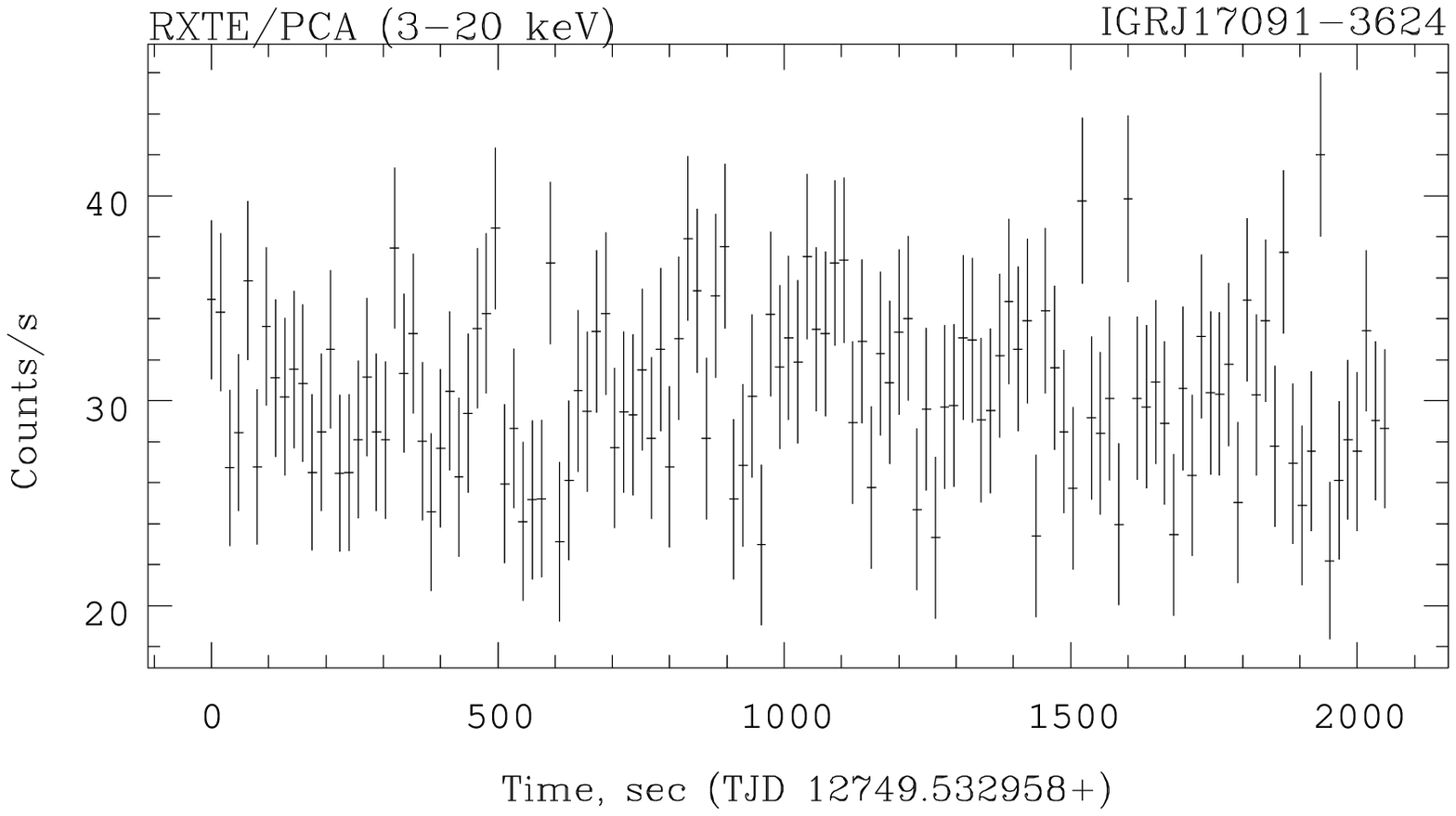}
\includegraphics[width=\columnwidth,bb=55 410 550 720,clip]{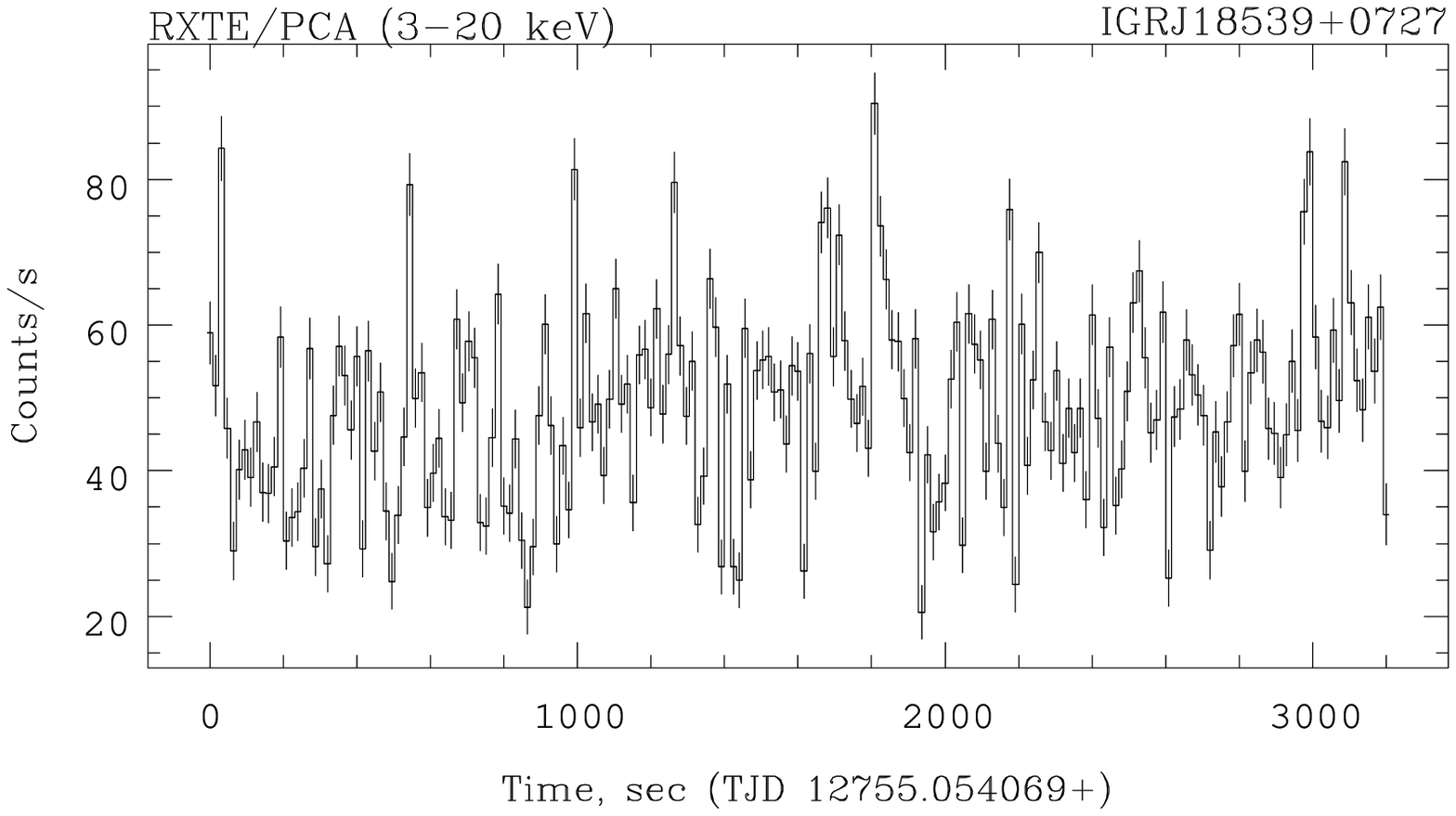}
}
\caption{Light curves of IGR J17091-3624 (left panel) and IGR J18539+0727 
(right panel), obtained by RXTE/PCA (3-20 keV). Every point corresponds 
to 16 sec time bin. Estimated contribution of galactic diffuse emission is
subtracted. \label{lcurve}}
\end{figure*}

X-ray sources IGR J17091-3624 and IGR J18539+0727 were observed by the
RXTE observatory (Bradt, Rothschild, Swank 1993)
 on April 20.5 and 26.1, 2003 respectively. Effective exposures
of performed observations were 2 and 3.2 ksec.

For RXTE data reduction we used standard programs of FTOOLS/LHEASOFT 5.2
package. For estimation of the PCA background we used ''L7\_240''-based
model. In order to increase our sensitivity to photons with energies higher
than 10-15 keV we used data of all layers of PCA detectors.
Because of weakness of sources they were not detected by HEXTE 
spectrometer (15-250 keV). For approximation of the sources spectra
we used $XSPEC$ package (http://xspec.gsfc.nasa.gov).

Spectrometer PCA of the RXTE observatory collects all photons from sky area
around 1 sq.deg without any possibility to distinguish between the
background photons and the photons from the source of our interest.
During observations of X-ray sources located close to the Galactic plane
the X-ray flux measured by PCA can contain significant contribution 
from ''background'' Galactic ridge emission. In the Galactic center
region at the Galactic latitude $\sim 2^\circ$ (where IGR J17091-3624
 is located) contribution of the Galactic ridge emission to the measured
PCA flux should be at the level of  $\sim0.5-0.6$ mCrab in 3-20 keV energy 
band (e.g. Valinia, Marshall 1998, Revnivtsev 2003).
At the same time, the flux detected from the source approximately equals to 
4 mCrab, or only 8 time higher than the ''background'' flux from the Galactic 
ridge. Because of that we should be very careful in our 
following data analysis. At the distance around $\sim$40$^\circ$
from the Galactic center and at the Galactic latitude $\sim3^\circ$, where  
IGR J18539+0727 is located, the contribution from the ridge emission
is significantly lower ($\sim$2-5\% of the source flux).

\begin{figure*}[t]
\hbox{
\includegraphics[width=\columnwidth]{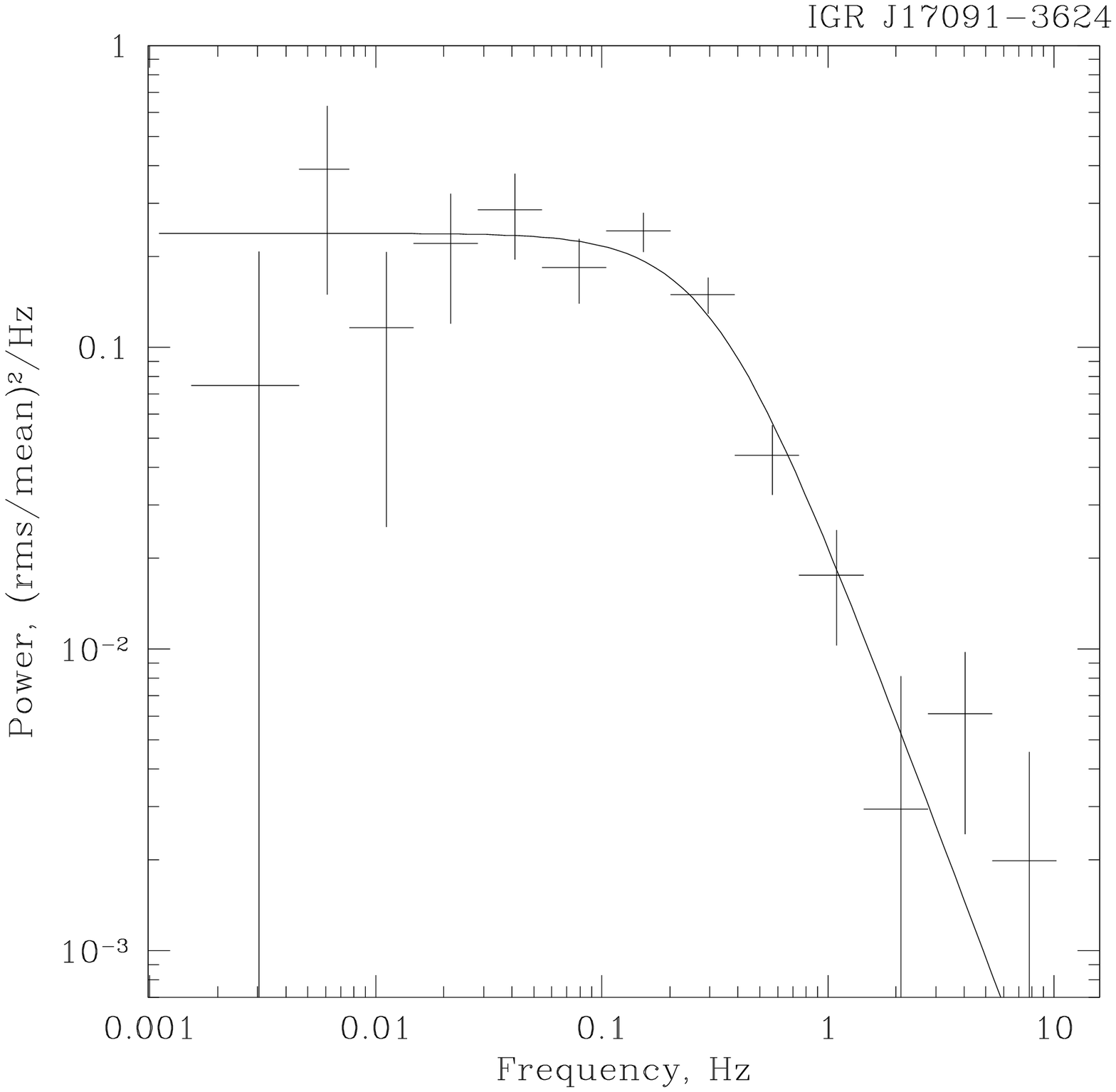}
\includegraphics[width=\columnwidth]{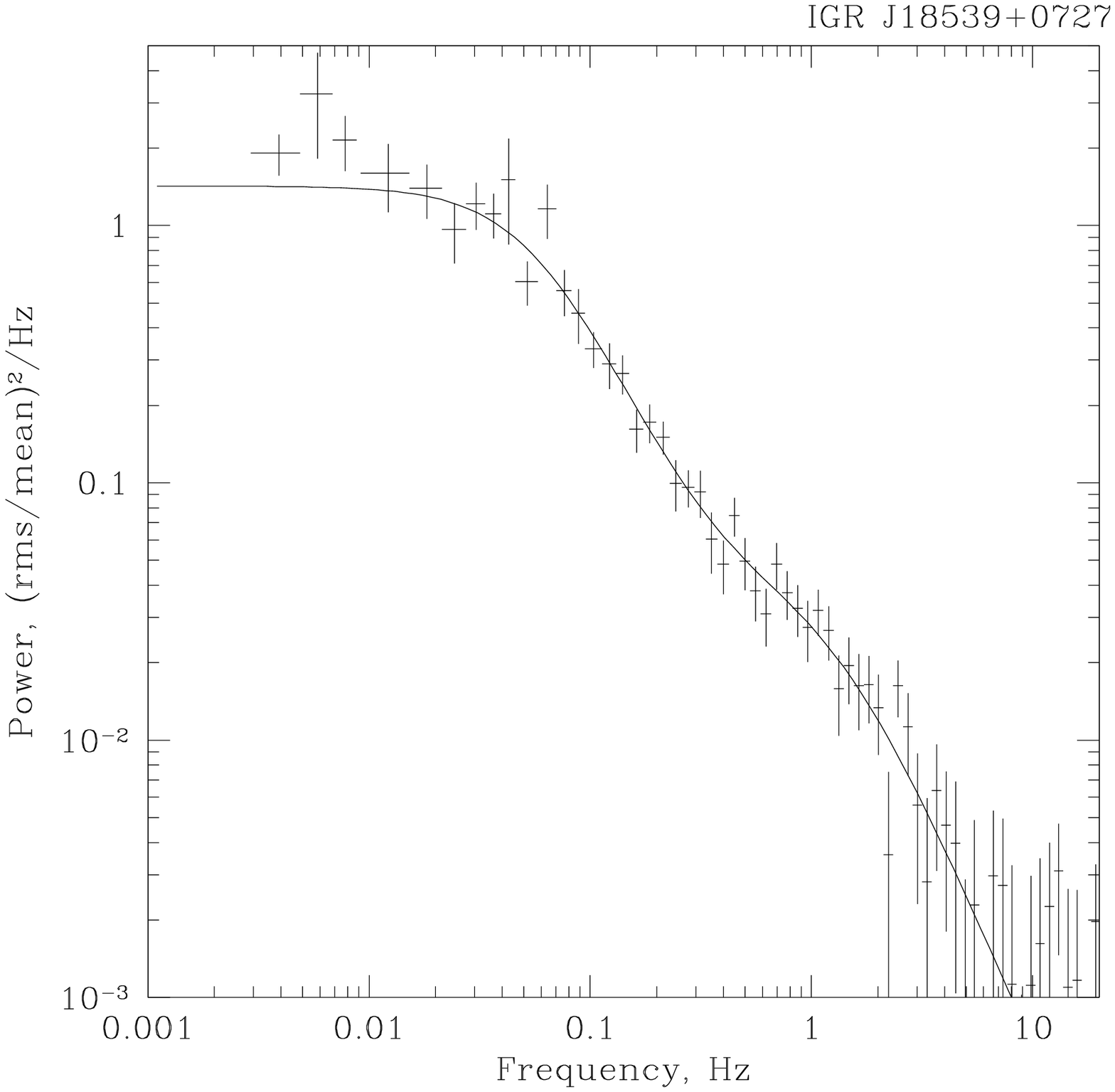}
}
\caption{Power spectra of  IGR J17091-3624 (left) and IGR J18539+0727 
(right). Models of power spectra approximation by band limited noise 
components are shown by solid lines. In the case of IGR J17091-3624 
one band limited noise component was used, in the case IGR J18539+0727 -- two.
\label{powersp}}

\end{figure*}

\begin{table*}[]
\caption{Best fit parameters of spectral approximations of IGR J17091-3624 and
IGR J18539+0727.}
\tabcolsep=9mm
\begin{tabular}{l|c|c}
\hline
\hline
Parameter&IGR J17091-3624$^a$&IGR J18539+0727\\
\hline
Neutral absorption $N_H L$, $10^{22}$ см$^{-2}$&$<1$&$1.5\pm0.4$\\
Photon index, $\Gamma$&$1.43\pm0.03$&$1.47\pm0.05$\\
Energy of the emission line, keV&$-^b$&$6.4\pm0.1$\\
Width of the line, keV&$-^b$&$<0.3$\\
Equivalent width of the line, EW, eV&$-^b$&$135\pm25$\\
Flux (3-25 keV), $10^{-10}$erg/s/cm$^2$&1.0&1.8\\
\hline
\end{tabular}

\vspace{2mm}
\begin{list}{}
\item $^a$ -- During approximation of the IGR J17091-3624 energy spectrum
the estimated contribution of Galactic ridge emission was subtracted.
Normalization of the ridge emission was taken to give 3-20 keV flux
at the level of $\sim 1.9\cdot 10^{-11}$ erg/s/cm$^2$/(in the field of 
view of PCA),  that is corresponds to result of Revnivtsev (2003)

\item $^b$ -- Influence of the Galactic ridge emission does not allow
us to state the presence of the emission line in the spectrum of the source
\end{list}
\end{table*}

\begin{figure*}
\hbox{
\includegraphics[width=\columnwidth]{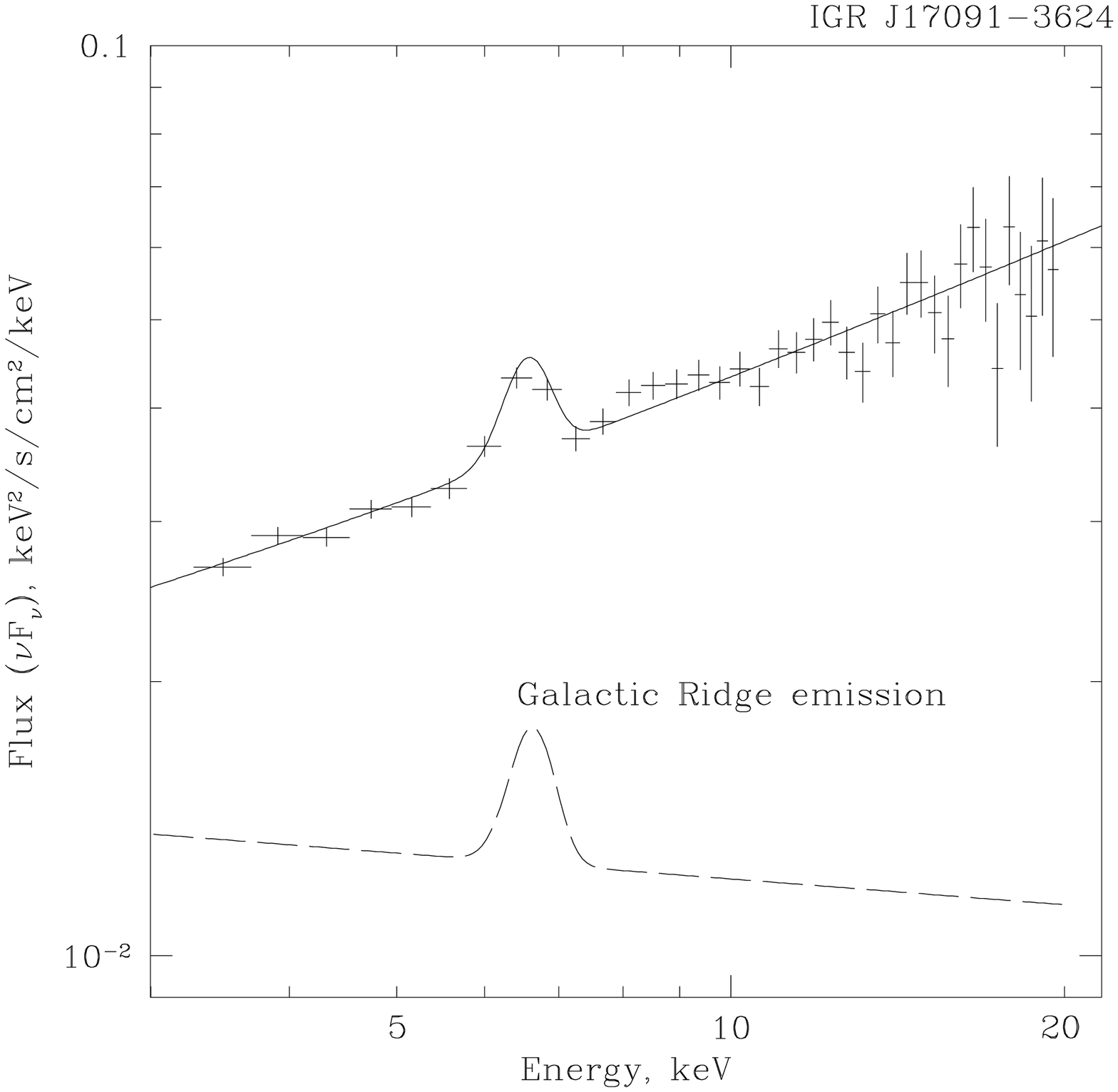}
\includegraphics[width=\columnwidth]{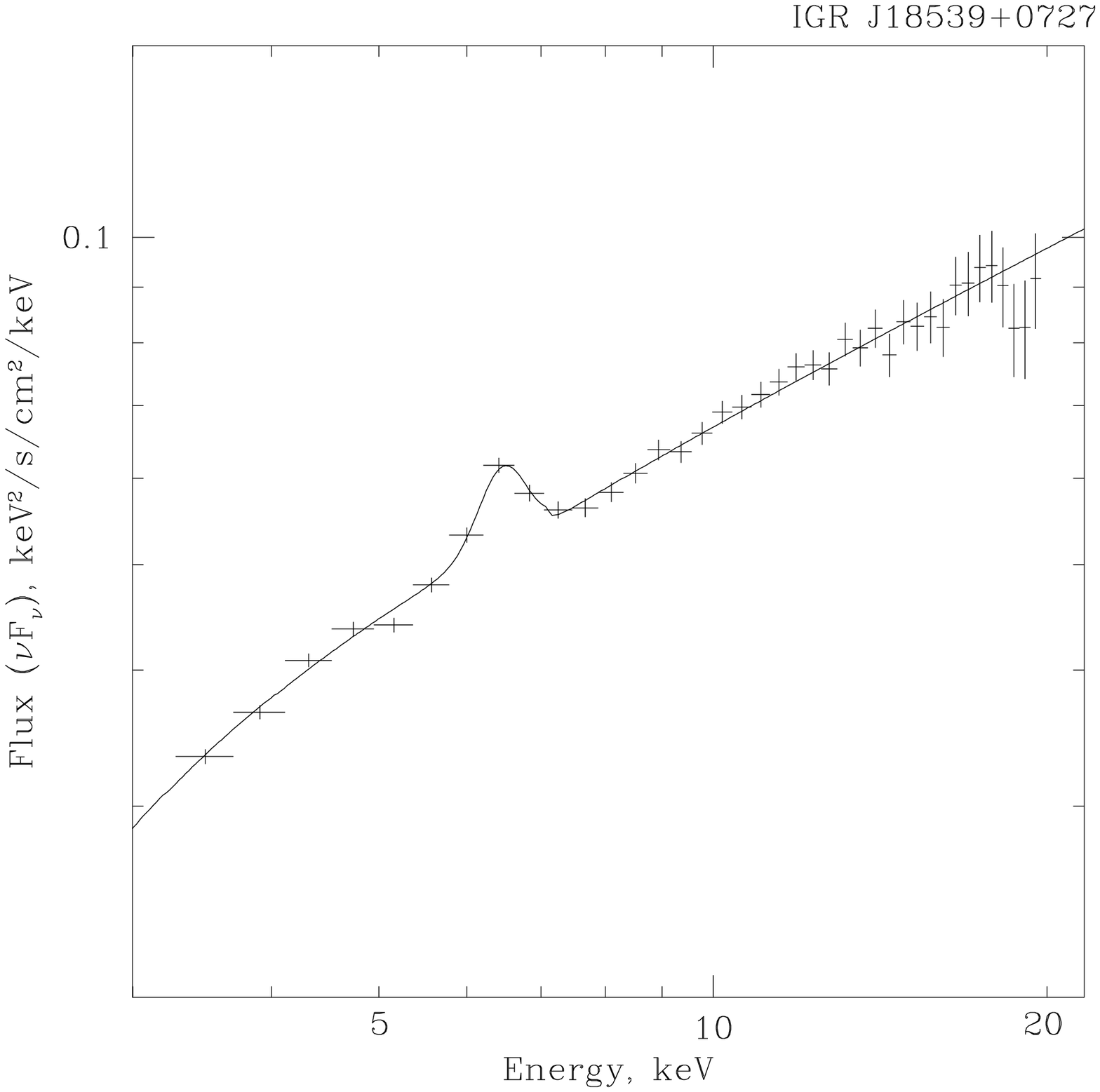}
}
\caption{ (left) Energy spectrum of IGR J17091-3624. Points with error bars
(1$\sigma$) represent the spectrum observed by RXTE/PCA, solid line 
shows the model of observed spectrum, that consist of source spectrum
and the spectrum of the galactic ridge emission (dashed line). 
(right) Energy spectrum of 
IGR J18539+0727. Points with error bars (1$\sigma$) represent the
observed spectrum, solid line shows the used model. Contribution
of the ridge emission is small.
 \label{spect}}
\end{figure*}

\subsection*{IGR J17091-3624}

Light curves of the source measured during observation on April 20, 2003
is shown in Fig. \ref{lcurve}(left). As it is seen from the figure
the source flux was relatively stable at the level of $\sim$ 4 mCrab 
(estimated contribution of the Galactic ridge emission is subtracted). 
 The fractional rms amplitude of its aperiodic variability equals
 to $27\pm3$\% in the frequency range 0.01-10 Hz.

No QPO-like features is seen on the power spectrum (Fig.\ref{powersp}(left)).
In general the power spectum
could be well approximated by standard model of band limited noise, 
which often is seen from black hole and neutron star binaries
in low/hard spectral state (e.g. Sunyaev, Revnivtsev 2003).
Approximation of the power spectrum with the model 
$dP\propto 1/(1+(f/f_0)^2) df$ gives the break frequency 
$f_0=0.31\pm0.04$ Hz.

In Fig.\ref{spect}(left) we present the energy spectrum of IGR J17091-3624,
averaged over all observation. Solid line shows the model used for
spectral approximation ($dN/dE\propto E^{-\Gamma}$). Dashed
line denotes the estimated contribution of the Galactic ridge emission 
to the spectrum collected by RXTE/PCA. Estimations based on measurements
of Revnivtsev (2003). As it is seen from the picture, the strong
influence of the ridge emission does not allow us to measure accurately
the parameters of the emission line, that could be present in the spectrum of
IGR J17091-3624. Moreover it is likely that all emission line photons 
has the Galactic ridge origin.
Best fit parameters of used spectral approximation are presented in Table 1.

\subsection*{IGR J18539+0727}

The light curve of IGR J18539+0727 in the 3-20 energy band is shown 
in Fig.\ref{lcurve}(right). The source is brighter then IGR J17091-3624 
and its strong 
aperiodic variability
is clearly seen. The points in the Figure are connected 
by histogram for
a clearness. 

The power spectrum of the source is presented in Fig.\ref{powersp}b.
Similar to IGR J17091-3624 the power spectrum of IGR J18539+0727
 could be well approximated by standard model 
of band limited noise. But the increase of sensitivity in a comparison
with previous case allowed us to detect the second component of the 
noise. Approximation of the power spectrum with the model
$dP_{1,2}\propto 1/(1+(f/f_{1,2})^2) df$ gives the break frequencies 
$f_1=(5.3\pm0.5)\cdot 
10^{-2}$ Hz and $f_2=1.3\pm0.2$ Hz.  The rms amplitude of aperiodic variability
contained in lower frequency component is $\sim 33$\%, in higher 
frequency component -- $\sim30$\%. Total source flux variability in 
frequency range 0.005-20 Hz equals to
$45\pm2$\%.

Energy spectrum of the source, averaged over whole observation, is shown
in Fig.\ref{spect}(right). Because IGR J18539+0727 is significantly further
from the Galactic center and by more than half a degree further from the 
Galactic plane in comparison with previous source, the contribution of 
the Galactic ridge emission to the
flux detected by RXTE/PCA is almost negligible. As it was noted before
its contribution does not exceed 2-5\% of the IGR J18539+0727 flux.

For the approximation of the spectrum of the source we used the same model - 
power law with neutral absorption. Apparent presence of the emission
line at the energy $\sim$6.4 keV probably caused by the reprocessing
the hard X-ray emission from the innermost parts of the accretion flow
by cold optically thick outer parts of the accretion disk (see e.g. 
Basko et al. 1974). However, the weakness of the source and low exposure
 of the used observation result in the low sensitivity of our spectral analysis
to the presence of the reflected/reprocessed component itself (continuum).
Upper limit on the possible amplitude of the reflected component (
model $pexrav$ of $XSPEC$ package; Magdziarz, Zdziaski 1995)
in the spectrum of  IGR J18539+0727 is $\Omega/2\pi <0.3$, where $\Omega$ is
the solid angle subtended by the reflector with respect to the source
of hard X-ray photons. Best fit parameters of a spectral approximation 
are presented in Table 1.

\section*{Discussion}

During several months from the start of operations of INTEGRAL observatory
several sources, which could be called weak hard transients, 
with roughly similar properties were discovered: 
IGR J17091-3624
(Kuulkers et al. 2003), IGR J18539+0727 (Lutovinov et al. 2003a),
IGR J18325-0756 (Lutovinov et al. 2003b), IGR17597-2201 
(Lutovinov et al. 2003c),
IGR J18483-0311 (Chernyakova et al. 2003). ``Hard'' in this case denotes
that transients, in spite of being quite weak, were still detected by 
INTEGRAL in a hard X-ray energy band 40-100 keV.

In this work we presented the results of spectral and timing analysis
of two sources - IGR J17091-3624 and IGR J18539+0727, that were observed
by RXTE satellite after the INTEGRAL discovery of them.

Spectra of these sources could be well approximated by power law model
 ($dN(E)\propto E^{-\alpha}dE$). Because of proximity of IGR J17091-3624 
to the Galactic center and, consequently, significant contribution 
of the Galactic ridge emission to the flux detected by RXTE/PCA,
we can not measure the parameters of possible fluorescent emission
line of the source. In the case of IGR J18539+0727 such analysis is
possible and we see the fluorescent line at energy $\sim$6.4 keV
(Fig.\ref{spect}, Table 1). Such fluorescent emission line is often observed
in spectra of X-ray binaries in low/hard spectral state. This
line likely originates as a result of reprocessing 
of hard X-ray emission of innermost parts of the accretion flow
by outer cold parts of optically thick accretion disk (see e.g
Basko et al. 1974, Gilfanov et al. 1999)

Both sources demonstrate strong flux variability on time scales from 
tenths to tens of seconds. Characteristics of energy spectra 
of sources and power spectra of their light curves allow
us to classify them as X-ray binary systems in low/hard spectral state.
Significant activity of IGR J17091-3624 in radio (Rupen et al. 2003)
indirectly confirms such classification (Fender, Hendry 2000)

For quite a long time the
parameters of power spectra of sources light curves are under discussion
in the view of possibility of determination of the nature of the 
compact object in the X-ray binary system (e.g. Wijnands \& van 
der Klis 1999, Sunyaev \& Revnivtsev 2000). It was demonstrated 
a number of times that the light curves of neutron star X-ray binaries
in the low spectral state typically contain higher frequency variability
than that of back hole X-ray binaries. Power spectra
of X-ray binaries allow one to apply strict criterion to distinguish
neutron star systems from back holes binaries (Sunyaev \& Revnivtsev 2000), 
however lower frequency band power spectra parameters do not allow
one to do so.

Investigations of X-ray binaries show that neutron star systems usually
do not demonstrate the lowest break frequency lower then $\sim$0.1 Hz
(Wijnands \& van der Klis 1999). In our case, the power spectrum of 
IGR J18539+0727 has the break frequency lower then that. Therefore
we can tentatively suggest that this source is a black hole candidate.

Parameters of power spectrum of IGR J17091-3624 do not allow us to do so,
however significant flux of the source in radio band observed by VLA
(Rupen et al. 2003) also hints that the source could be a black hole
candidate (Fender et al. 2001). In spite of that, for real determination
of the nature of compact objects in these systems one need more
observations, especially in the optical spectral band, that would
allow to identify optical companions of X-ray sources and measure
the mass functions of the binaries.

\bigskip
{\it

Authors thank Jean Swank and and RXTE planning team for RXTE
 TOO observations of IGR J17091-3624 and IGR J18539+0727.
Research has made use of data obtained 
through the High Energy Astrophysics Science Archive Research Center 
Online Service, provided by the NASA/Goddard Space Flight Center.

Work is partially supported by RFBR grant 02-02-17347, grant of 
Minpromnauka (NSH-2083.2003.2) and program of Russian 
Academy of Sciences ``Unstationary phenomena in astronomy''.

}

\section*{References}

\indent

Basko M., Sunyaev R., Titarchuk L., Astron. Astroph., {\bf 31}, 249 (1974)

Bradt, Rothschild, Swank J., 
Astron. Astrophys. Suppl. Ser.  {\bf 97}, 355 (1993)

Chernyakova M., Lutovinov A., Capitanio F., et. al.,  Astron. Telegram
ATEL {\bf 157} (2003)

Fender R., Hendry M., MNRAS, {\bf 317}, 1 (2000)

Fender R., MNRAS, {\bf 322}, 31 (2001)

Gilfanov M., Churazov E., Revnivtsev M., Astron. Astroph. {\bf 352}, 182 (1999)

in't Zand J., J. Heise J., Lowes P., et. al.,  Astron. Telegram
ATEL {\bf 160} (2003)

Kuulkers E., Lutovinov A., Parmar A, et. al., Astron. Telegram
ATEL {\bf 149} (2003)

Lutovinov A., Rodriguez J., Produit N., et. al., Astron. Telegram
ATEL {\bf 151} (2003a)

Lutovinov A., Shaw S., Foschini L., et. al., Astron. Telegram
ATEL {\bf 154} (2003b)

Lutovinov A., Walter R., Belanger G., et. al., Astron. Telegram
ATEL {\bf 155} (2003c)

Magdziarz P., Zdziarski A., MNRAS, {\bf 273}, 837 (1995)

Revnivtsev M.,  Gilfanov M., Churazov E., Sunyaev R.,  Astron. Telegram
ATEL {\bf 150} (2003)

Revnivtsev M., Astron. Astrophys. in press (2003), (astro-ph/0304351)

Rupen M., Mioduszewski A., Dhawan V.,  Astron. Telegram
ATEL {\bf 152} (2003)

Sunyaev R., Revnivtsev M., Astron. Astroph. {\bf 358}, 617 (2000)

Valinia A., Marshall F., Astroph. J., {\bf 505}, 134 (1998)

Wijnands R., van der Klis M., Astroph.J., {\bf 514}, 939 (1999)

\end{document}